\documentclass[journal,10pt]{IEEEtran}

\usepackage{cite}
\usepackage{graphicx}
\usepackage{amsmath}
\usepackage{array}
\usepackage{mdwmath}
\usepackage{amssymb}
\usepackage{mdwtab}
\usepackage{stfloats}
\usepackage[tight,footnotesize]{subfigure}
\usepackage{amsmath,amsthm}
\usepackage{threeparttable}
\usepackage{color}
\usepackage{url}
\usepackage{algpseudocode}
\usepackage{algorithm}
\usepackage{multicol}
\usepackage{amssymb}
\usepackage{epstopdf}
\usepackage{makecell}

\algrenewcommand{\algorithmicrequire}{\textbf{Input:}}
\algrenewcommand{\algorithmicensure}{\textbf{Output:}}

\newtheorem{remark}{Remark}
\newtheorem{lemma}{Lemma}

\begin{document}
\title{Transmissive RIS Transmitter Enabled Spatial Modulation for MIMO Systems}
\author{
Xusheng Zhu, Qingqing Wu, \IEEEmembership{Senior Member, IEEE}, and Wen Chen, \IEEEmembership{Senior Member, IEEE}

\thanks{
(\emph{Corresponding author: Qingqing Wu.})}
\thanks{X. Zhu, Q. Wu, and W. Chen are with the Department of Electronic Engineering, Shanghai Jiao Tong University, Shanghai 200240, China (e-mail: xushengzhu@sjtu.edu.cn; qingqingwu@sjtu.edu.cn; wenchen@sjtu.edu.cn).}
}

\maketitle
\begin{abstract}
In this paper, we propose a novel transmissive reconfigurable intelligent surface (TRIS) transmitter-enabled spatial modulation (SM) multiple-input multiple-output (MIMO) system.
In the transmission phase, a column-wise activation strategy is implemented for the TRIS panel, where the specific column elements are activated per time slot. Concurrently, the receiver employs the maximum likelihood detection technique.
Based on this, for the transmit signals, we derive the closed-form expressions for the upper bounds of the average bit error probability (ABEP) of the proposed scheme from different perspectives, employing both vector-based and element-based approaches. Furthermore, we provide the asymptotic closed-form expressions for the ABEP of the TRIS-SM scheme, as well as the diversity gain.
To improve the performance of the proposed TRIS-SM system, we optimize ABEP with a fixed data rate. Additionally, we provide lower bounds to simplify the computational complexity of improved TRIS-SM scheme.
The Monte Carlo simulation method is used to validate the theoretical derivations exhaustively.
The results demonstrate that the proposed TRIS-SM scheme can achieve better ABEP performance compared to the conventional SM scheme. Furthermore, the improved TRIS-SM scheme outperforms the TRIS-SM scheme in terms of reliability.
\end{abstract}

\section {Introduction}
With the explosive growth in data traffic and the number of wireless connections, the power consumption of base stations and the cost of deploying them has increased exponentially. To cope with this challenge, multiple-input multiple-output (MIMO) techniques leverage space-time multiplexing and diversity techniques to improve the effectiveness and reliability of the system\cite{big2007mimo}. Due to expensive radio-frequency (RF) links behind each antenna, it can significantly increase the cost of deployment. Nonetheless, fifth-generation (5G) technology employs massive MIMO technology to further meet data transmission rates and reliability via hybrid analog-digital beamforming \cite{and2014what}.
By 2030, it is expected that the data rate of sixth-generation (6G) technology will reach 1 Tb/s, the spectrum efficiency will reach 100 bps/Hz, and the latency will be less than 1 ms \cite{dang2020what}. These requirements are crucial for integrated sensing and communication, immersive communication, etc. In addition to meeting various emerging application scenarios, 6G technology also needs to satisfy individual voice requirements, which requires exploring new multi-antenna technologies to reduce power consumption and costs of wireless networks.

\begin{table*}[t]
\centering
\caption{\small{Notations in this paper}}
\begin{tabular}{|c|c|c|c|}
\hline Notations & Definitions & Notations & Definitions \\
\hline$N_{{t}}$ & The number of transmit antennas & $N_{{r}}$ & The number of receive antennas \\
\hline$L$ & The number of transmissive RIS units & $\mathbb{C}^{m\times n}$ & The space of $m\times n$ matrics\\
\hline $N$ & \thead{The number of columns of \\ transmissive RIS units} & $L_N$ & \thead{The number of rows of\\ transmissive RIS units} \\
\hline $\mathbb{C}^{m\times n}$ & The space of $m\times n$ matrics & $[\cdot]^T$ & The transpose operation \\
\hline $M$ & The number of modulation orders  & {$\|\cdot\|$} &2-norm operation\\
\hline diag\{$\cdot$\} & The diagonal matrix operation of the vector & $\otimes$ & The Kronecker product operation \\
\hline $[\cdot]^H$ & The conjugate transpose operation & det($\cdot$) & The determinant operation for matrices \\
\hline $\frac{d^{m}}{dt^{m}}\left\{F\right\}$ & \thead{The $m$-th order derivative of the \\ $F$ function with respect to $t$}  & $(\cdot)!$ & The factorial operation \\
\hline ${N\choose n}$ & \thead{The number of combinations of \\ $n$ chosen from $N$}  &$\mathcal{CN}(\cdot,\cdot)$ &  The complex Gaussian distribution\\
\hline $\Pr(\cdot)$ & The probability of an event happening & $\Re\{\cdot\}$ & The real part operation for complex value \\
\hline $\mathcal{N}(\cdot,\cdot)$ &  The real Gaussian distribution & $Q(\cdot)$ & $Q(x)=\int_x^\infty\frac{1}{\sqrt{2\pi}}\exp(-\frac{t^2}{2})dt$ \\
\hline$E[\cdot]$ & The expectation operation on a variable & $Var[\cdot]$ & The variance of a variable \\
\hline$\Im\{\cdot\}$ & The imaginary part operation for complex value  & $|\cdot|$ & The absolute value operation \\
\hline$(\cdot)!!$ & The double factorial operation  & $\|\cdot\|_F$ & The Frobenius norm operation \\
\hline$(\cdot)^*$ & The conjugate operation on complex value & $\mathcal{D}$ & Diversity gain  \\
\hline$P_b(\mathbf{x}\to\hat{\mathbf x}|\mathbf{H})$ &{CPEP} &$P_b(\mathbf{x}\to\hat{\mathbf x})$ & {UPEP} \\
\hline $P_b([n,m]\to[\hat{n},\hat m]|\mathbf{H})$ & CPEP & $P_b([n,m]\to[\hat{n},\hat m])$ & UPEP \\
\hline $f_X(\cdot)$ & The probability density function (PDF) on $X$ & $\sim$ & ``Distributed as" \\
\hline $\sin(\cdot)$ & Sine function & ${\rm tan}(x)$ & Tangent function \\
\hline
\end{tabular}
\end{table*}

Reconfigurable intelligent surfaces (RIS) is regarded as a promising and economical support technology for future wireless \cite{wu2024int}.
More specifically, RIS comprises an extensive array of passive, cost-effective elements, arranged in a two-dimensional planar configuration. Within this array, each RIS element is independently tunable, allowing for precise control over its amplitude, phase, and frequency through a dedicated RIS controller \cite{pan2021rec}. This capability circumvents the need for complex encoding and decoding processes.
As a result, each element has a high degree of flexibility by adjusting the corresponding coefficient\cite{huang2019rec}.
With this in mind, the RIS enables the reconfigurability of the wireless channel by regulating the phase of the elements to enhance useful signals or suppress interfering signals \cite{liu2021rec}.
Currently, most RIS-related work is based on reflective RIS-assisted communication techniques \cite{wu2019int}. Typically, reflective RIS is deployed in the channel to enhance the signal or extend the area coverage.
Recently, \cite{liu2023trans} proposed a transmitter architecture based on transmissive RIS (TRIS), which is a very promising transceiver technology.
An important reason for this is that the structure of TRIS-based transmitter is quite simple compared to the conventional transmitter \cite{li2023tow}. Specifically, the TRIS transmitter consists of a horn-fed antenna, a controller, and a TRIS panel \cite{tang2023tran}.
Since the incident and transmissive signals of the TRIS transmitter are located on both sides of the TRIS, interference between electromagnetic signals can be effectively avoided \cite{li2023tow}. In addition, TRIS has higher aperture efficiency and larger operating bandwidth \cite{bai200high}.
Based on the above considerations, the application of TRIS to multi-antenna transmitter design is a prospective technology for future wireless communications.


In addition to the RIS technique, spatial modulation (SM) based on the MIMO architecture is a promising technique, especially in terms of improving system spectral efficiency and reducing power consumption\cite{mesleh2008saptial}.
Specifically, the SM technique combines conventional $M$-ary phase shift keying/quadrature amplitude modulation (PSK/QAM) with antenna-indexed modulation, thereby adding one degree of freedom to the modulation \cite{wen2019a}.
Since SM is equipped with an RF link on the implementation side, it activates only one corresponding antenna in each transmission time slot according to the input bit stream \cite{li2021single}. In contrast to the MIMO technique, SM effectively avoids problems such as signal interference and synchronization among multiple antennas \cite{{ren2014spatial}}.
To facilitate the implementation of SM, \cite{zhu2021on} conducted real-world measurements of the SM technology in both indoor corridor and lecture room environments with channel measurement techniques.
In each environment, both line-of-sight (LoS) and non-LoS (NLoS) scenarios were analyzed in the practical measurements.
In \cite{mes2015qua}, the authors proposed the quadrature SM (QSM) method by preprocessing the transmit signals. Within this, the spatial domain constellation symbols are expanded into in-phase and quadrature components and activate both transmit antennas simultaneously.
Hence, the QSM not only improves the spectral efficiency, but also avoids the interference between the transmit signals.
In addition, the authors in \cite{youn2010gen} proposed the generalized SM (GSM) system, where the SM is generalized from activating one antenna to activating multiple transmitting antennas within the same time slot. It is worth mentioning that GSM can achieve higher spectral efficiency with fewer antennas.
To improve the average bit error probability (ABEP) performance of the SM system, the authors proposed the adaptive SM (ASM) scheme in \cite{yang2011adap} by fixing the average data rate.
For millimeter-wave (mmWave) band signal transmission, the reliability of the signal is difficult to guarantee if the transmission is carried out the SM due to the severe path loss.
In views of this, \cite{ding2017spg} and \cite{zhu2023qud} extend SM to the mmWave band and present schemes for spatial scattering modulation (SSM) and quadrature SSM, respectively.

Enlightened by their potential advantages, the RIS-aided SM systems have been investigated in a large number of communication scenes \cite{basar2020rec,zhu2024on,li2023int,luo2021spatial,ma2020lar,hbo2021ris,li2024novel,san2022int,zhu2023ris}.
To be specific, the authors of \cite{basar2020rec} considered the RIS-assisted receive SM scenario, where maximum likelihood (ML) detection and greedy detection algorithms are applied at the receiver side, together with the corresponding analytical ABEP expressions.
Intending to enhance the reliability of RIS-assisted receive SM, \cite{li2024novel} proposed the novel RIS-assisted receive SM scheme, where the alternating active and passive beamforming optimization method is employed to concentrate the energy of the target receive antenna while reducing the energy leakage to the other receive antennas.
Compared to the receive SM, it is more common to adopt the SM technique at the transmitter side.
For this reason, \cite{zhu2024on} conducted the investigation of RIS-assisted transmit SM for downlink communication system, where the Gaussian-Chebyshev quadrature approach is exploited to derive the ABEP closed-form expression.
To improve the spectral efficiency of RIS-assisted SM, \cite{ma2020lar} combines the transmitting and receiving antenna indices for joint modulation, where the reflected beam is shaped by RIS taking both the transmit SM and the receive SM into account.
From RIS-assisted transmit and receive SM perspectives, \cite{luo2021spatial} investigated the formation of an optimization problem by minimizing the symbol error rate, which is addressed by adopting the penalty-alternative method.
By taking advantage of the information carrying capacity of SM and the reconfigurability of RIS to the environment, \cite{li2023int} studied the potential applications of RIS-SM in relay networks, wireless sensing, and cognitive radio.
Besides, the framework of RIS-assisted SM-based ambient backscattering communication was presented in \cite{hbo2021ris}, where the tag device can implement ambient backscattering to harness the energy of RF waves in the environment as well as SM techniques.
Additionally, \cite{san2022int} proposed the RIS-assisted transceiver-based QSM scheme in which the transmitter and receiver select a set of antenna indexes for transmitting and receiving the real and imaginary parts of the complex signal, respectively. It is worth mentioning that the RIS is logically divided into two parts for reflecting these portions of the signal.
For RIS-assisted SM transmission in the mmWave band, \cite{zhu2023ris} proposed the RIS-SSM scheme, where the RIS is deployed closer to the transmitter and further away from the receiver. Consequently, these two sub-channels are considered for both LoS and NLoS communication scenarios, respectively.

Up to now, all the studies related to RIS and SM are based on reflective RIS-assisted communication, which improves the ABEP performance of SM through the deployment strategy of reflective RIS.
Besides, it can achieve better system performance by deploying RIS closer to the transceiver in accordance with \cite{wu2019int}.
Additionally, since TRIS has the advantage of a simple structure, it can greatly alleviate the problem of high power consumption and the high cost of traditional transceivers, which makes it a promising architecture.
For these reasons, we propose the TRIS transmitter-enabled SM-based MIMO scheme.

\begin{itemize}
\item We propose a novel TRIS transceiver-enabled SM-MIMO architectural system. To the best of our knowledge, work based on TRIS-enabled SMs has not been reported in the open literature. In the proposed TRIS-SM scheme, the TRIS activates a corresponding column of TRIS elements at each instant based on the input bit information. Then, the activated TRIS elements radiate the PSK/QAM signals in the symbol domain into free space.
\item At the receiver, we employ an ML detector to retrieve the transmitted signal.
Based on this, we first derive the conditional pairwise error probability (CPEP) of the proposed scheme by adopting the vector-based (VB) approach for the transmit signals.
Subsequently, leveraging the CPEP and the moment-generating function (MGF), we obtain the closed-form expression for the unconditional pairwise error probability (UPEP).
To offer another idea of derivation, we obtain the UPEP closed-form expression for the transmit signals by adopting the element-based (EB) approach.
In order to gain further valuable insights, we also provide the UPEP asymptotic expressions and diversity gains for the proposed TRIS-SM scheme.
Finally, the joint upper bound expression for ABEP is given by combining the derived UPEP and Hamming distance.
\item Aiming to improve the reliability for the proposed scheme, we form an optimization problem of minimizing ABEP. By effective transformation, we obtain a target problem of maximizing the minimum Euclidean distance with the constraint of fixing the amount of data in each time slot. Afterward, to reduce the complexity of the improved TRIS-SM scheme, we simplify it and propose a lower bound for the improved TRIS-SM scheme.
\item Extensive Monte Carlo simulation results are utilized to verify the correctness of the theoretically derived results. It is shown that the ABEP results obtained using both VB and EB approaches are identical. Additionally, we find that the proposed TRIS-SM scheme is to obtain better ABEP performance than the conventional SM. Furthermore, the improved TRIS-SM scheme is also superior to the TRIS-SM scheme in terms of reliability.
\end{itemize}

The remainder of this paper can be organized as follows.
In Section II, the proposed system model for TRIS-enabled SM is described.
The ABEP performance of the proposed TRIS-SM system is provided in Section III.
In Section IV, the improved TRIS-SM scheme is presented.
The simulation results, analytical results and corresponding discussions are given in Section V.
Finally, Section VI concludes the whole paper.
Note that the Notations in this paper are summarized in Table I.

\section {System Model}

\begin{figure}[t]
  \centering
  \includegraphics[width=8.5cm]{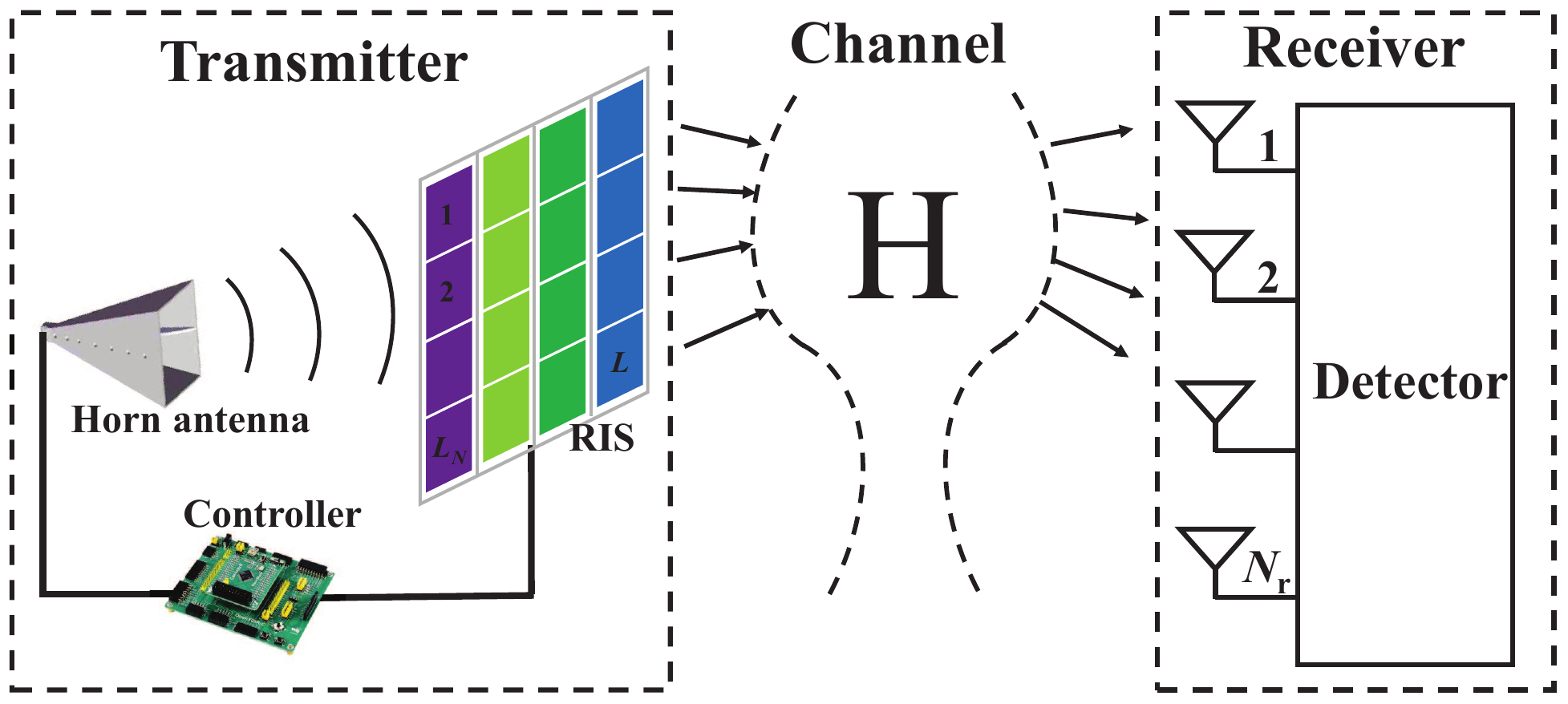}
  \caption{\small{System model.}}\label{sys}
\end{figure}

We consider the TRIS-SM scheme for MIMO system as depicted in Fig. \ref{sys}, where the transmitter consists of a feedback horn antenna, a TRIS, and $N_r$ receive antennas.
In particular, the TRIS composed of $L$ low-cost passive units is connected to the feedback horn antenna through the controller.
Besides, the transmissive coefficient of the TRIS can be characterized as
\begin{equation}\label{eq1}
\mathbf{f} = [f_{1},\cdots,f_{l},\cdots,f_{L}]^T\in\mathbb{C}^{L\times 1},
\end{equation}
where we assume that the TRIS is completely transmissive and there is no signal reflection.  Also, the TRIS is made up of a uniform planar array (UPA). Since the implementation of element-level control of the TRIS is costly,
We adopt the column-control approach \cite{chen2020angle} with respect to TRIS plane.
Notice that the $L$ transmissive units consist of $N$ column elements, and each column is composed of $L_N$ transmissive units, such that $L_N={L}/{N}$.
At this point, the controller only needs to regulate $N$ effective units to achieve the regulation of all the transmissive units.

Without loss of generality, we make the vector $\mathbf{e} = [e_1,\cdots,e_n,\cdots, e_N]^T$ denote the column-control signal of TRIS. For the control signal of the $n$-th column, we have
$
e_n = f_{(n-1)L_N+1}=f_{(n-1)L_N+2}=\cdots = f_{nL_N}=\beta_ne^{j\phi_n},
$
where $\beta_n$ and $\phi_n$ denote the magnitude and phase shared by the $n$-th column unit of the TRIS, respectively.
In this paper, the TRIS phase is set to a constant value $\phi_0=0$ \cite{can2020rec}.
Meanwhile, the amplitude of the TRIS is regulated by the controller with on-off ($\beta_n=1$ for on, $\beta_n=0$ for off).
Taking this into account, the phase and amplitude adjustment mechanism of the TRIS implemented in this paper can significantly reduce the control complexity and increase the practicality.
In this manner, (\ref{eq1}) can be further expressed as
\begin{equation}
\mathbf{e}_n = [\underbrace{0,\cdots,0}_{n-1},\underbrace{1}_{n-{\rm th \ coordinate}},\underbrace{0,\cdots, 0}_{N-n}]^T.
\end{equation}

For the proposed TRIS-SM scheme, the input information consists of two parts, i.e., the symbol information and the additional spatial domain information.
Specifically, the symbol bit information is mapped to constellation points from $M$-ary phase
shift keying/quadrature amplitude modulation (PSK/QAM) and then radiated through the horn antenna to the TRIS panel, while the other part of the information is mapped to activate the corresponding column elements of the TRIS.
Here, the information of the $m$-th constellation point of the symbol domain and the $n$-th column of the spatial domain activating the transmissive element can be described as
\begin{equation}
\mathbf{x}_{m,n}=s_m\mathbf{e}_n \in \mathbb{C}^{1\times N},
\end{equation}
where $s_m \in\{s_1,\cdots,s_m,\cdots,s_M\}$ denotes the symbol from $M$-ary PSK/QAM diagram of constellation.
For this purpose, the transmit symbols of the joint spatial and symbol domains at each transmission time slot can be written as
\begin{equation}
\mathbf{x} =[\mathbf{x}_{1,n}^T,\cdots, \mathbf{x}_{m,n}^T, \cdots, \mathbf{x}_{M,n}^T]^T\in\mathbb{C}^{L\times 1}.
\end{equation}
On the basis of this, the received signal vector $\mathbf{y}$ can be obtained as
\begin{equation}\label{channel1}
\mathbf{y}=\mathbf{Hx}+\mathbf{n},
\end{equation}
where
$\mathbf{H}\in\mathbb{C}^{N_r\times L}$ stands for the independent and identically distributed (i.i.d.) wireless fading channel whose elements are distributed with a complex Gaussian distribution with zero mean and unit variance, $\mathbf{x}$ belongs to the TRIS-SM signal set $\mathcal{S}$,
and $\mathbf{n}$ denotes the complex additive white Gaussian noise (AWGN) vector that has covariance matrix $N_0I_{N_r}$.
It is worth noting that the signal $\mathbf{x}$ can be given by
\begin{equation}
\mathcal{S}=\left\{s_1\mathbf{e}_1,s_2\mathbf{e}_2,\cdots,s_M\mathbf{e}_N\right\}.
\end{equation}
Assume that the input symbols output with the same probability, we employ the optimal ML detector in terms of average error performance at the receiver side to retrieve the transmit signal, which can be formulated as
\begin{subequations}\label{ml}
\begin{align}
\hat {\mathbf{x}}&=\arg\max\limits_{\mathbf{x} \in\mathcal{S}}p_{\mathbf{y}}(\mathbf{y}|\mathbf{x},\mathbf{H})\\
&=\arg\min\limits_{\mathbf{x} \in\mathcal{S}}\|\mathbf{y}-\mathbf{Hx}\|^2,
\end{align}
\end{subequations}
where $\hat{\mathbf{x}}$ denotes the detected signal,
$p_{\mathbf {y}}(\cdot)$ stands for the PDF of $\mathbf{y}$, and $\mathcal{S}$ represents the set of all legitimate transmit symbols in the spatial and symbol domains.

\section{Performance Analysis}
In this section, we derive the UPEP closed-form expressions depending on the form of the transmit symbol elements by employing both VB and EB methods, respectively. From this, we also provide the asymptotic UPEP expression and ABEP union upper bound expression for the TRIS-SM scheme.
Furthermore, we provide the diversity gain of the proposed scheme.
\subsection{VB Method to UPEP Derivation}
According to (\ref{ml}), the {CPEP} expression between the transmitted signal $\mathbf{x}$ and the detected signal $\hat{\mathbf{x}}$ can be derived as
\begin{equation}\label{cpepchar1}
\begin{aligned}
P_b(\mathbf{x}\to\hat{\mathbf x}|\mathbf{H})
=&\Pr(\|\mathbf{y}-\mathbf{Hx}\|^2>\|\mathbf{y}-\mathbf{H}\hat{\mathbf{x}}\|^2)\\
=&\Pr(\|\mathbf{n}\|^2>\|\mathbf{n}+\mathbf{H}(\mathbf{x}-\hat{\mathbf{x}})\|^2)\\
=&\Pr(\|\mathbf{n}\|^2-\|\mathbf{n}+\mathbf{H}(\mathbf{x}-\hat{\mathbf{x}})\|^2>0)\\
=&\Pr(G>0),
\end{aligned}
\end{equation}
where $G = \|\mathbf{n}\|^2-\|\mathbf{n}+\mathbf{H}(\mathbf{x}-\hat{\mathbf{x}})\|^2$.
To facilitate the representation, we can set
\begin{equation}
\mathbf{w} = \mathbf{n}+\mathbf{H}(\mathbf{x}-\hat{\mathbf{x}}).
\end{equation}
In this case, we can rewrite $G$ as
\begin{equation}\label{g1}
G = \|\mathbf{n}\|^2-\|\mathbf{w}\|^2=\mathbf{u}^H\mathbf{Gu},
\end{equation}
where $\mathbf{u}=
\begin{bmatrix}
\mathbf{n}^T&\mathbf{w}^T
\end{bmatrix}^T
$ and $\mathbf{G} = {\rm diag}\{\mathbf{I}_{N_r},-\mathbf{I}_{N_r}\}$.
Next, we adopt {\bf Lemma 1} to obtain the distribution of the random vector $\mathbf{u}$.
\begin{lemma}
The random vector $\mathbf{u}=\begin{bmatrix}\mathbf{n}^T&\mathbf{w}^T\end{bmatrix}^T$ follows the circularly symmetric complex Gaussian random distribution, and its covariance can be given by
\begin{equation}
\mathbf{K_u}=
\begin{bmatrix}
N_0\mathbf{I}_{N_r} & N_0\mathbf{I}_{N_r}\\
N_0\mathbf{I}_{N_r} & (N_0+\|\mathbf{x}-\hat{\mathbf{x}}\|^2)\mathbf{I}_{n_r}
\end{bmatrix}.
\end{equation}

Proof:
Let us make the matrices $\mathbf{X}=\mathbf{I}_{N_r}\otimes \mathbf{x}^T$ and $\hat{\mathbf{X}}=\mathbf{I}_{N_r}\otimes \hat{\mathbf{x}}^T$, where $\mathbf{X}\in\mathbb{C}^{N_r\times L}$ and $\hat{\mathbf{X}}\in\mathbb{C}^{N_r\times L}$.
And then, we define $\mathbf{h}_{n}= [h_{n,1},h_{n,2},\cdots,h_{n,N_r}]^T$.
Based on this, we have
\begin{equation}
\mathbf{Hx} = \mathbf{Xh}_{n}, \ \ \mathbf{H\hat x} = \mathbf{\hat X h}_{n}.
\end{equation}
Subsequently, we can respectively rewrite $\mathbf{w}$ and $\mathbf{u}$ as
\begin{equation}
\mathbf{w} = (\mathbf{X-\hat X})\mathbf{h}_{n}+\mathbf{n},
\end{equation}
and
\begin{equation}
\mathbf{u} =
\begin{bmatrix}
\mathbf{n} \\ \mathbf{w}
\end{bmatrix}
=\underbrace{
\begin{bmatrix}
\boldsymbol{0}_{N_r\times L} & \mathbf{I}_{N_r}\\
\mathbf{X-\hat X} & \mathbf{I}_{N_r}
\end{bmatrix}}_{\triangleq \mathbf{A}}
\underbrace{
\begin{bmatrix}
\mathbf{h}_{n}\\
\mathbf{n}
\end{bmatrix}
}_{\triangleq \mathbf{v}}.
\end{equation}
It can be observed that the random vector $\mathbf{v}$ follows the circularly-symmetric complex Gaussian distribution. The corresponding covariance can be given by
\begin{equation}
\mathbf{K_v}={\rm diag}\{N_0\mathbf{I}_{N_r},\mathbf{I}_{N_r}\}.
\end{equation}
In views of this, the covariance of variable $\mathbf{u}$ can be given by
\begin{equation}
\begin{aligned}
\mathbf{K_u}&=\mathbf{AK_vA}^H\\&=
\begin{bmatrix}
N_0\mathbf{I}_{N_r} & N_0\mathbf{I}_{N_r}\\
N_0\mathbf{I}_{N_r}&(\mathbf{X-\hat X})(\mathbf{X-\hat X})^H+N_0\mathbf{I}_{N_r}
\end{bmatrix}.
\end{aligned}
\end{equation}
By substituting $(\mathbf{X}-\hat{\mathbf{X}})(\mathbf{X}-\hat{\mathbf{X}})^H=\|\mathbf{x}-\hat{\mathbf{x}}\|^2\mathbf{I}_{N_r}$, the proof is completed.$\hfill\blacksquare$
\end{lemma}
Based on {\bf Lemma 1},  the variable $G$ in (\ref{g1}) is a complex Gaussian random variable subject to Hermitian quadratic form. For this reason, the corresponding moment-generating function can be expressed as \cite{turin1960the}
\begin{equation}\label{mgf1}
\begin{aligned}
MGF_G(t)&=\frac{1}{{\rm det}(\mathbf{I}_{2N_r}-t \mathbf{K_uG})}\\
&=\frac{1}{{\rm det}(\mathbf{I}_2-t\tilde{\mathbf{K}}_{\mathbf u}\otimes \mathbf{I}_{N_r})}\\
&=\frac{1}{{\rm det}(\mathbf{I}_2-t\tilde{\mathbf{K}}_{\mathbf u})^{N_r}},
\end{aligned}
\end{equation}
where $MGF_X\left(t\right)=\int_0^\infty\exp(tx)\exp(-x)dx=\frac{1}{1-t}$ denotes the moment generating function (MGF) with variables of $X$ and coefficients of $t$, and
\begin{equation}
\tilde{\mathbf{K}}_{\mathbf{u}}=
\begin{bmatrix}
N_0 & -N_0\\
N_0 & -N_0-\|\mathbf{x}-\hat{\mathbf{x}}\|^2
\end{bmatrix}.
\end{equation}
Following that, we move on to addressing the eigenvalues of matrix $\tilde{\mathbf{K}}_{\mathbf{u}}$ as
\begin{equation}\label{det1}
\begin{aligned}
&{\rm det}(\lambda-\tilde{\mathbf{K}}_{\mathbf{u}})^2
={\rm det}\left(
\begin{matrix}
\lambda - N_0 & N_0\\
-N_0 & \lambda+N_0+\|\mathbf{x}-\hat{\mathbf{x}}\|^2
\end{matrix}
\right)^2=0.
\end{aligned}
\end{equation}
After some algebraic manipulation, (\ref{det1}) can be recast as
\begin{equation}
\lambda^2+\lambda\|\mathbf{x}-\hat{\mathbf{x}}\|^2-N_0\|\mathbf{x}-\hat{\mathbf{x}}\|^2=0.
\end{equation}
With the aid of the rule for addressing quadratic equations, we respectively attain the eigenvalues $\lambda_1$ and $\lambda_2$ as
\begin{equation}\label{lamb1}
\lambda_1 = \frac{-\|\mathbf{x}-\hat{\mathbf{x}}\|^2+\sqrt{\|\mathbf{x}-\hat{\mathbf{x}}\|^4+4N_0\|\mathbf{x}-\hat{\mathbf{x}}\|^2}}{2},
\end{equation}
\begin{equation}\label{lamb2}
\lambda_2 = \frac{-\|\mathbf{x}-\hat{\mathbf{x}}\|^2-\sqrt{\|\mathbf{x}-\hat{\mathbf{x}}\|^4+4N_0\|\mathbf{x}-\hat{\mathbf{x}}\|^2}}{2}.
\end{equation}
Depending on the eigenvalues (\ref{lamb1}) and (\ref{lamb2}) of $G$, we can reformulate (\ref{mgf1}) as
\begin{equation}
MGF_G(t)=\frac{1}{(1-t\lambda_1)^{N_r}(1-t\lambda_2)^{N_r}}.
\end{equation}
By applying the inversion theorem \cite{Pe1951not} and the residual theorem \cite{ah1979com}, the (\ref{cpepchar1}) can be reconstructed as
\begin{equation}\label{cpepchar2}
\begin{aligned}
&P_b(\mathbf{x}\to\tilde{\mathbf{x}})\\
&=\frac{-1}{(N_r-1)!}\frac{d^{N_r-1}}{dt^{N_r-1}}\left\{
\left(t-\frac{1}{\lambda_2}\right)^{N_r}\frac{MGF_G(t)}{t}
\right\}_{\bigg|t=\frac{1}{\lambda_2}}\\
&=\frac{(-1)^{N_r-1}}{(N_r-1)!\lambda_2^{N_r}}\frac{d^{N_r-1}}{dt^{N_r-1}}\left\{
\frac{1}{t(1-\lambda_1t)^{N_r}}
\right\}_{\bigg|t=\frac{1}{\lambda_2}}.
\end{aligned}
\end{equation}
After utilizing the Leibniz differentiation rule \cite{olv1980app}, we can further characterize the derivation result in (\ref{cpepchar2}) as
\begin{equation}\label{cpepchar3}
\begin{aligned}
&\frac{d^{N_r-1}}{dt^{N_r-1}}\left\{
\frac{1}{t(1-\lambda_1t)^{N_r}}
\right\}\\
&=\sum_{n_r=0}^{N_r-1}{N_r-1 \choose n_r}
\underbrace{\frac{d^{N_r-1-n_r}}{dt^{N_r-1-n_r}}
\left\{\frac{1}{t}
\right\}}_{\triangleq B_1}
\underbrace{
\frac{d^{n_r}}{dt^{n_r}}\left\{
\frac{1}{(1-\lambda_1t)^{N_r}}
\right\}}_{\triangleq B_2}.
\end{aligned}
\end{equation}
In this respect, we respectively address the derivatives of $B_1$ and $B_2$ of (\ref{cpepchar3}) as
\begin{equation}\label{cpepchar4}
B_1
=
\frac{(-1)^{N_r-1-n_r}(N_r-1-n_r)!}{t^{N_r-n_r}},
\end{equation}
\begin{equation}\label{cpepchar5}
B_2
=\frac{\lambda_1^{n_r}\Pi_{j=0}^{{n_r}-1}(N_r+j)}{(1-\lambda_1t)^{N_r+n_r}}.
\end{equation}
By substituting (\ref{cpepchar4}) and (\ref{cpepchar5}) into (\ref{cpepchar3}), the $\frac{d^{N_r-1}}{dt^{N_r-1}}\left\{
\frac{1}{t(1-\lambda_1t)^{N_r}}
\right\}_{|t=\frac{1}{\lambda_2}}$ in (\ref{cpepchar2}) can be obtained as
\begin{equation}\label{cpepchar6}
\begin{aligned}
&\frac{d^{N_r-1}}{dt^{N_r-1}}\left\{
\frac{1}{t(1-\lambda_1t)^{N_r}}
\right\}_{|t=\frac{1}{\lambda_2}}\\
&=\sum_{n_r=0}^{N_r-1}\frac{(N_r+n_r-1)!}{n_r!}\frac{-\lambda_1^{n_r}\lambda_2^{2N_r}}{(\lambda_1-\lambda_2)^{N_r+n_r}}.
\end{aligned}
\end{equation}
In views of (\ref{cpepchar2}) and (\ref{cpepchar6}), the UPEP expression of the proposed TRIS-SM scheme can be written as
\begin{equation}\label{cpepchar7}
\begin{aligned}
&P_b(\mathbf{x}\to\hat{\mathbf{x}})
\\=&
\frac{(-1)^{N_r-1}}{(N_r-1)!\lambda_2^{N_r}}
\sum_{{n_r}=0}^{N_r-1}\frac{(N_r+{n_r}-1)!}{{n_r}!}\times\frac{-\lambda_1^{n_r}\lambda_2^{2N_r}}{(\lambda_1-\lambda_2)^{N_r+k}}\\
=&\left(\frac{-\lambda_2}{\lambda_1-\lambda_2}\right)^{N_r}\sum_{{n_r}=0}^{N_r-1}{N_r+{n_r}-1\choose {n_r}}\left(\frac{\lambda_1}{\lambda_1-\lambda_2}\right)^{n_r}.
\end{aligned}
\end{equation}

\subsection{EB Method to UPEP Derivation}
To provide another thought of the derivation, we split the transmit signal to map to the channel selection and symbol domain information regarding the RIS column elements separately.
To be specific, (\ref{channel1}) can be equivalently expressed as
\begin{equation}\label{channel2}
\mathbf{y}=\mathbf{H}_{n}s_m+\mathbf{n},
\end{equation}
where $\mathbf{H}_n\in\mathbb{C}^{N_r\times L_N}$ denotes the channel from the $n$-th column element of the RIS to the $N_r$ receive antennas.
Since the different $\mathbf{H}_{n,l}$ are i.i.d. to $l$ and follow $\mathcal{CN}(0,\mathbf{I}_{N_r})$, we can obtain the distribution of $\mathbf{h}_n \in \mathbb{C}^{N_r \times 1} $ as
\begin{equation}\label{hncn}
\mathbf{h}_n=\sum\nolimits_{l=1}^{L_N}\mathbf{H}_{n,l}\sim \mathcal{CN}(0,L_N\mathbf{I}_{N_r}).
\end{equation}
Based on this, the CPEP can be written as
\begin{equation}\label{cpep2}
\begin{aligned}
&P_b([n,m]\to[\hat{n},\hat m]|\mathbf{H})\\
=&\Pr \left(\|\mathbf{y}-\mathbf{h}_{n}s_m\|^2>\|\mathbf{y}-\mathbf{h}_{\hat n}s_{\hat m}\|^2\right)\\
=&\Pr\left(\|\mathbf{n}\|^2>\|\mathbf{y}-\mathbf{h}_{\hat n}s_{\hat m}\|^2\right)\\
=&\Pr \left(\|\mathbf{n}\|^2>\|\mathbf{y}\|^2+\|\mathbf{h}_{\hat n}s_{\hat m}\|^2-2\Re\{\mathbf{h}_{\hat n}s_{\hat m}(\mathbf{h}_{n}s_m+\mathbf{n})\}\right)\\
=&\Pr \left(\|\mathbf{n}\|^2>\|\mathbf{h}_{n}s_m+\mathbf{n}\|^2+\|\mathbf{h}_{\hat n}s_{\hat m}\|^2\right.\\&\left.-2\Re\{\mathbf{h}_{n}\mathbf{h}_{\hat n}s_ms_{\hat m}+\mathbf{h}_{\hat n}s_{\hat m}\mathbf{n}\}\right)\\
=& \Pr \left(\|\mathbf{n}\|^2>\|\mathbf{h}_{n}s_m\|^2+\|\mathbf{n}\|^2+2\Re\{\mathbf{h}_{n}s_m\mathbf{n}_0\}\right.\\&\left.+\|\mathbf{h}_{\hat n}s_{\hat m}\|^2-2\Re\{\mathbf{h}_{n}\mathbf{h}_{\hat n}s_ms_{\hat m}+\mathbf{h}_{\hat n}s_{\hat m}\mathbf{n}\}\right).
\end{aligned}
\end{equation}
After some mathematical manipulation, we can further obtain the (\ref{cpep2}) as
\begin{equation}\label{cpep3}
\begin{aligned}
&P_b([n,m]\to[\hat{n},\hat m]|\mathbf{H})\\
=&\Pr (0>\|\mathbf{h}_{n}s_m\|^2+2\Re\{\mathbf{h}_{n}s_m\mathbf{n}\}\\&+\|\mathbf{h}_{\hat n}s_{\hat m}\|^2-2\Re\{\mathbf{h}_{n}\mathbf{h}_{\hat n}s_ms_{\hat m}+\mathbf{h}_{\hat n}s_{\hat m}\mathbf{n}\})\\
=&\Pr (-\|(\mathbf{h}_{n}s_m-\mathbf{h}_{\hat n}s_{\hat m})\|^2-2\Re\{(\mathbf{h}_{n}s_m-\mathbf{h}_{\hat n}s_{\hat m})\mathbf{n}\}>0)\\
=& \Pr\left(G>0\right),
\end{aligned}
\end{equation}
where $G = -\|(\mathbf{h}_{n}s_m-\mathbf{h}_{\hat n}s_{\hat m})\|^2-2\Re\{(\mathbf{h}_{n}s_m-\mathbf{h}_{\hat n}s_{\hat m})\mathbf{n}\}$ variable is distributed with the Gaussian distribution.
Accordingly, it follows that $G\sim\mathcal{N}(\mu_G,\sigma_G^2)$, where the corresponding expectation and variance are $\mu_G=-\|(\mathbf{h}_{n}s_m-\mathbf{h}_{\hat n}s_{\hat m})\|^2$ and $\sigma_G^2=2N_0\|\mathbf{h}_{n}s_m-\mathbf{h}_{\hat n}s_{\hat m}\|^2$, respectively.
Towards this end, (\ref{cpep3}) can be equivalent to
\begin{equation}\label{cpep1}
\begin{aligned}
&P_b([n,m]\to[\hat{n},\hat m]|\mathbf{H}) \\&= Q\left(\sqrt{\frac{\mu_G^2}{\sigma_G^2}}\right)=Q\left(\sqrt{\frac{\left\|\mathbf{h}_{n}s_m-\mathbf{h}_{\hat n}s_{\hat m}\right\|^2}{2N_0}}\right)\\&=Q\left(\sqrt{\frac{\sum_{n_r=1}^{N_r}\left|{h}_{n_r,n}s_m-{h}_{ n_r,\hat n}s_{\hat m}\right|^2}{2N_0}}\right).
\end{aligned}
\end{equation}
It is easy to see that (\ref{cpep1}) does not satisfy the closed form of the expression due to the presence of Q-function. In this regard, we define $\kappa_{n_r}=\left|{h}_{n_r,n}s_m-{h}_{ n_r,\hat n}s_{\hat m}\right|^2$ and use {\bf Lemma 2} to statistically average it.

\begin{lemma}
For $\kappa_{n_r}=\left|{h}_{n_r,n}s_m-{h}_{ n_r,\hat n}s_{\hat m}\right|^2$, we treat the symbol domain information as invariant parameters and carry out the expectation operation for the i.i.d. Rayleigh channel only. In this case, we have
\begin{equation}
\bar\kappa_{n_r}=\left\{
\begin{aligned}
&|s_m-s_{\hat m}|^2L_N, \ \ \ \hat n_t = n_t\\
&(|s_m|^2+|s_{\hat m}|^2)L_N, \hat n_t \neq n_t\\
\end{aligned}\right.
\end{equation}

Proof:
Taking an expectation operation on $\kappa_{n_r}$, we have
\begin{equation}\label{kapp1}
\bar\kappa_{n_r}
= E\left[\left|{h}_{n_r,n_t}s_m-{h}_{ n_r,\hat n_t}s_{\hat m}\right|^2\right].
\end{equation}
To facilitate the subsequent derivation, we split (\ref{kapp1}) into the form of adding the real and imaginary parts of the information. In parallel, we make use of the expectation addition operation law for independent signals, which yields
\begin{equation}\label{kapp1_2}
\begin{aligned}
\bar\kappa_{n_r}
=& E\left[\left|\Re\{{h}_{n_r,n_t}s_m-{h}_{ n_r,\hat n_t}s_{\hat m}\}\right|^2\right]\\
&+E\left[\left|\Im\{{h}_{n_r,n_t}s_m-{h}_{ n_r,\hat n_t}s_{\hat m}\}\right|^2\right].
\end{aligned}
\end{equation}
In the following, we discuss two cases based on whether the spatial domain signals are correctly detected or not \cite{zhu2023ssmp}.

For the RIS column elements signal is correctly detected, we have

\begin{equation}\label{kapp2}
\begin{aligned}
\bar\kappa_{n_r}
=& E\left[\left|\Re\{{h}_{n_r,n_t}(s_m-s_{\hat m})\}\right|^2\right]\\
&+E\left[\left|\Im\{{h}_{n_r,n_t}(s_m-s_{\hat m})\}\right|^2\right].
\end{aligned}
\end{equation}
Since the spatial and symbol domain signals are independent of each other, we can further transform (\ref{kapp2}) via the independent signal expectation multiplication relation into
\begin{equation}\label{kapp3}
\begin{aligned}
\bar\kappa_{n_r}
=& E\left[\left|\Re\{{h}_{n_r,n_t}\}\right|^2\right]|s_m-s_{\hat m}|^2\\
&+E\left[\left|\Im\{{h}_{n_r,n_t}\right|^2\}\right]|s_m-s_{\hat m}|^2.
\end{aligned}
\end{equation}
By merging the real and imaginary spatial domain signal energies of (\ref{kapp3}), we obtain
\begin{equation}\label{kapp4}
\begin{aligned}
\bar\kappa_{n_r}
=& E\left[\left|{h}_{n_r,n_t}\right|^2\right]|s_m-s_{\hat m}|^2.
\end{aligned}
\end{equation}
Based on $E\left[\left|{h}_{n_r,n_t}\right|^2\right] = \left|E\left[{h}_{n_r,n_t}\right]\right|^2+Var[{h}_{n_r,n_t}]$, we first need to obtain $|E\left[{h}_{n_r,n_t}\right]$ and $Var[{h}_{n_r,n_t}]$.
To address this issue, we resort to (\ref{hncn}). Thus, (\ref{kapp4}) can be recast as
\begin{equation}\label{kapp4_5}
\begin{aligned}
\bar\kappa_{n_r}
=& |s_m-s_{\hat m}|^2L_N.
\end{aligned}
\end{equation}

For the case of incorrect detection of the RIS column element signals, the spatial domain transmit and detect signals are independent of each other. Hence, we can further rewrite (\ref{kapp1_2}) as
\begin{equation}
\begin{aligned}
\bar\kappa_{n_r}
=& E\left[\left|\Re\{{h}_{n_r,n_t}\}\right|^2\right]|s_m|^2+E\left[\left|\Re\{{h}_{ n_r,\hat n_t}\}\right|^2\right]|s_{\hat m}|^2\\
&+E\left[\left|\Im\{{h}_{n_r,n_t}\}\right|^2\right]|s_m|^2+E\left[\left|\Im\{{h}_{ n_r,\hat n_t}\}\right|^2\right]|s_{\hat m}|^2.
\end{aligned}
\end{equation}
Further, we can merge the same symbol field entries to arrive at
\begin{equation}
\begin{aligned}
\bar\kappa_{n_r}
=& E\left[\left|{h}_{n_r,n_t}\right|^2\right]|s_m|^2+E\left[\left|{h}_{ n_r,\hat n_t}\right|^2\right]|s_{\hat m}|^2.
\end{aligned}
\end{equation}
By leveraging (\ref{hncn}), we can get
\begin{equation}\label{kapp5}
\bar\kappa_{n_r}
=|s_m|^2L_N+|s_{\hat m}|^2L_N.
\end{equation}
In combining the two cases (\ref{kapp4_5}) and (\ref{kapp5}), we complete the proof.
$\hfill\blacksquare$
\end{lemma}

\begin{remark}
In $\kappa_{n_r}$, we find that $h_{n_r,n}$ is the channel with respect to $L_N$ mutually independent transmission elements.
Meanwhile, Lemma 1 contains $L_N$ in the variance.
Therefore, the relationship between the random variable A and the mean B can be characterized as
\begin{equation}
\kappa_{n_r} = \bar \kappa_{n_r} x,
\end{equation}
where $x$ stands for the central chi-square distribution with 2 degrees of freedom and a variance of 0.5. Thus, the PDF can be given as
\begin{equation}
f_X(x)=\exp(-x).
\end{equation}
\end{remark}

\subsubsection{Single Receive Antenna $(N_r=1)$}
In this case, the UPEP can be written as
\begin{equation}\label{pb2}
P_b([n,m]\to[\hat{n},\hat m]) = \int_0^\infty Q\left(\sqrt{\frac{\rho\bar\kappa x}{2}}\right)f_X(x)dx.
\end{equation}
With the aid of \cite{zhu2024per}, the Q-function can be equivalently expressed as
\begin{equation}\label{qfun}
Q(x)=\frac{1}{\pi}\int_0^{\frac{\pi}{2}}\exp\left(-\frac{x^2}{2\sin^2\theta}\right)d\theta.
\end{equation}
Substituting (\ref{qfun}) into (\ref{pb2}), we can rewrite UPEP as
\begin{equation}
P_b([n,m]\!\to\![\hat{n},\hat m])
\!=\! \frac{1}{\pi}\int_0^\infty\int_0^{\frac{\pi}{2}} \exp\left({\frac{-\rho\bar\kappa x}{4\sin^2\theta}}\right)f_X(x)d\theta dx.
\end{equation}
By exchanging the order of integration of $\theta$ and $x$, we yield
\begin{equation}\label{pb3}
P_b([n,m]\!\to\![\hat{n},\hat m])
\!=\! \frac{1}{\pi}\int_0^{\frac{\pi}{2}}\int_0^\infty \exp\left({\frac{-\rho\bar\kappa x}{4\sin^2\theta}}\right)f_X(x) dx d\theta.
\end{equation}
Fortunately, the inner integrals in (\ref{pb3}) can be handled such that it can be simplified as
\begin{equation}\label{pb4}
P_b([n,m]\!\to\![\hat{n},\hat m])
\!=\! \frac{1}{\pi}\int_0^{\frac{\pi}{2}}
MGF_X\left(\frac{-\rho\bar\kappa}{4\sin^2\theta}\right)
d\theta,
\end{equation}
Considering this, (\ref{pb4}) can be evaluated as
\begin{equation}
P_b([n,m]\!\to\![\hat{n},\hat m])
= \frac{1}{\pi}\int_0^{\frac{\pi}{2}}
\frac{4\sin^2\theta}{4\sin^2\theta+{\rho\bar\kappa}}
d\theta.
\end{equation}
After some substitution operations, we have
\begin{equation}\label{pb5}
P_b([n,m]\!\to\![\hat{n},\hat m])
=\frac{1}{2}-\frac{1}{\pi}\int_0^{\frac{\pi}{2}}
\frac{{\rho\bar\kappa}}{4\sin^2\theta+{\rho\bar\kappa}}
d\theta.
\end{equation}
By making use of \cite{jeff2007tab}, we can rewrite (\ref{pb5}) as
\begin{equation}\label{pb6}
\begin{aligned}
&P_b([n,m]\!\to\![\hat{n},\hat m])\\
=&\frac{1}{2}-{\frac{\rho\bar\kappa}{4\pi}}\sqrt{\frac{16}{{\rho\bar\kappa}(4+{\rho\bar\kappa})}}\tan^{-1}\left(\sqrt{\frac{4+{\rho\bar\kappa}}{{\rho\bar\kappa}}}\tan\theta\right)\bigg|_{0}^{
\frac{\pi}{2}}.
\end{aligned}
\end{equation}
On this basis, we can derive the UPEP as
\begin{equation}
P_b([n,m]\!\to\![\hat{n},\hat m])
=\frac{1}{2}\left(1-\sqrt{\frac{{\rho\bar\kappa}}{{4}+{\rho\bar\kappa}}}\right).
\end{equation}

\subsubsection{Multiple Receive Antennas $(N_r>1)$}
In this case, the channel between the $n$-th column element vector of the TRIS and the $N_r$ receive antennas can be characterized as
\begin{equation}\label{nrpb1}
\begin{aligned}
\mathbf{f_X}(\mathbf{x})=f_{X_1,\cdots,X_{n_r},\cdots, X_{N_r}}
(x_1,\cdots,x_{n_r},\cdots, x_{N_r}).
\end{aligned}
\end{equation}
Considering both (\ref{cpep1}) and (\ref{nrpb1}), the UPEP of the TRIS-SM scheme with multiple receive antennas can be calculated as
\begin{equation}\label{nrpb2}
\begin{aligned}
P_b([n,m]\!\to\![\hat{n},\hat m])
=&\int_0^\infty Q\left(\sqrt{\frac{\rho\bar\kappa x}{2}}\right)\mathbf{f_X({x})}d\mathbf{x}.
\end{aligned}
\end{equation}
Since the channel between the transmitter and receiver is subject to i.i.d. Rayleigh distribution, we can further rewrite (\ref{nrpb1}) as
\begin{equation}\label{nrpb3}
\begin{aligned}
\mathbf{f_X({x})}&=f_{X_1}(x_1)\cdots f_{X_{n_r}}(x_{n_r})\cdots f_{X_{N_r}}(x_{N_r})\\
&=\prod\limits_{n_r=1}^{N_r}f_{n_r}(x_{n_r}).
\end{aligned}
\end{equation}
Taking (\ref{nrpb3}) into (\ref{nrpb2}), we can rewrite UPEP as
\begin{equation}\label{nrpb4}
\begin{aligned}
P_b([n,m]\!\to\![\hat{n},\hat m])
=&\underbrace{ \int_0^\infty\int_0^\infty\cdots\int_0^\infty}_{N_r-{\rm fold}}\prod\limits_{n_r=1}^{N_r} Q\left(\sqrt{\frac{\rho\bar\kappa x_{n_r}}{2}}\right)\\&\times f_{n_r}(x_{n_r})dx_1dx_2\cdots dx_{N_r}.
\end{aligned}
\end{equation}
After an equivalent transformation of the Q-function, we can further evaluate (\ref{nrpb4}) as
\begin{equation}\label{nrpb5}
\begin{aligned}
&P_b([n,m]\!\to\![\hat{n},\hat m])\\
=&\underbrace{ \int_0^\infty\int_0^\infty\cdots\int_0^\infty}_{N_r-{\rm fold}} \frac{1}{\pi}\int_0^{\frac{\pi}{2}}\prod\limits_{n_r=1}^{N_r}\exp\left(\frac{-\rho\bar\kappa x_{n_r}}{4\sin^2\theta}\right)\\&\times f_{n_r}(x_{n_r})d\theta dx_1dx_2\cdots dx_{N_r}.
\end{aligned}
\end{equation}
Since it is difficult to address (\ref{nrpb5}) directly, we convert it by swapping the order of integrals of $\mathbf{x}$ and $\theta$ to
\begin{equation}\label{nrpb6}
\begin{aligned}
&P_b([n,m]\!\to\![\hat{n},\hat m])\\
=&\frac{1}{\pi}\int_0^{\frac{\pi}{2}}\underbrace{ \int_0^\infty\int_0^\infty\cdots\int_0^\infty}_{N_r-{\rm fold}} \prod\limits_{n_r=1}^{N_r}\exp\left(\frac{-\rho\bar\kappa x_{n_r}}{4\sin^2\theta}\right)\\&\times f_{n_r}(x_{n_r}) dx_1dx_2\cdots dx_{N_r}d\theta.
\end{aligned}
\end{equation}
By the definition of the {MGF}, (\ref{nrpb6}) can be further recast as
\begin{equation}\label{nrpb7}
P_b([n,m]\!\to\![\hat{n},\hat m])
=\frac{1}{\pi}\int_0^{\frac{\pi}{2}}\prod\limits_{n_r=1}^{N_r}
 MGF_{X_{n_r}}\left(\frac{-\rho\bar\kappa}{4\sin^2\theta}\right) d\theta.
\end{equation}
As the signals arriving from the transmitter to the different receive antennas satisfy the i.i.d. condition, thus (\ref{nrpb7}) can be re-expressed as
\begin{equation}\label{nrpb78}
\begin{aligned}
P_b([n,m]\!\to\![\hat{n},\hat m])
=&\frac{1}{\pi}\int_0^{\frac{\pi}{2}}\left(\frac{4\sin^2\theta}{4\sin^2\theta+{\rho\bar\kappa}}\right)^{N_r} d\theta.
\end{aligned}
\end{equation}
In dependence on \cite{sim2001digital} and (\ref{pb6}), the UPEP of the considered system can be given by
\begin{equation}\label{nrpb8}
P_b([n,m]\!\to\![\hat{n},\hat m])
=\mu_\alpha^{N_r}\sum\limits_{{n_r}=0}^{N_r-1}{{N_r}-1+n_r\choose {n_r}}\left(1-\mu_\alpha\right)^{n_r},
\end{equation}
where $\mu_\alpha = \frac{1}{2}\left(1-\sqrt{\frac{{\rho\bar\kappa}}{{4}+{\rho\bar\kappa}}}\right)$.

\subsubsection{Asymptotic UPEP expression}
Based on the UPEP expression obtained above, we approximate UPEP to obtain the asymptotic expression in the high signal-to-noise ratio (SNR) region as
\begin{equation}
P_{\rm asy}([n,m]\!\to\![\hat{n},\hat m])=\lim\limits_{\rho \to \infty}P_b([n,m]\!\to\![\hat{n},\hat m]).
\end{equation}
It can be observed that it is hard to derive the asymptotic UPEP directly from (\ref{cpepchar7}) through VB method and from (\ref{nrpb8}) through EB method. To tackle this issue, we resort to (\ref{nrpb78}) as
\begin{equation}\label{asyeq1}
P_{\rm asy}([n,m]\!\to\![\hat{n},\hat m])
=\lim\limits_{\rho \to \infty}\frac{1}{\pi}\int_0^{\frac{\pi}{2}}\left(\frac{{4}\sin^2\theta}{{4}\sin^2\theta+{\rho\bar\kappa}}\right)^{N_r} d\theta.
\end{equation}
When the SNR is large, the denominator part $4\sin^2\theta$ in (\ref{asyeq1}) becomes quite small compared to $\rho\bar\kappa$. By neglecting $4\sin^2\theta$, we can rewrite (\ref{asyeq1}) as
\begin{equation}\label{asyeq2}
P_{\rm asy}([n,m]\!\to\![\hat{n},\hat m])
=\frac{1}{\pi}\int_0^{\frac{\pi}{2}}\left(\frac{4\sin^2\theta}{{\rho\bar\kappa}}\right)^{N_r} d\theta.
\end{equation}
By separating the variables of (\ref{asyeq2}), we get
\begin{equation}\label{asyeq3}
P_{\rm asy}([n,m]\!\to\![\hat{n},\hat m])
=\frac{1}{\pi}\left(\frac{4}{{\rho\bar\kappa}}\right)^{N_r} \int_0^{\frac{\pi}{2}}\left({\sin\theta}\right)^{2N_r} d\theta.
\end{equation}
Finally, we can take advantage of the Wallis formula \cite{jeff2007tab} to simplify (\ref{asyeq3}) to
\begin{equation}\label{asyeq4}
P_{\rm asy}([n,m]\!\to\![\hat{n},\hat m])=\frac{1}{2}\left(\frac{4}{{\rho\bar\kappa}}\right)^{N_r} \frac{(2N_r-1)!!}{(2N_r)!!}.
\end{equation}
\subsubsection{Diversity Gain}
In accordance with \cite{zheng2003diver}, the diversity gain of the proposed scheme can be characterized as
\begin{equation}\label{diver1}
\mathcal{D}=-\lim\limits_{\rho \to \infty}\frac{\log_2P_{\rm asy}(n\to\hat{n})}{\log_2\rho}.
\end{equation}
Substituting (\ref{asyeq3}) into (\ref{diver1}), we have
\begin{equation}\label{diver2}
\mathcal{D}
=-\lim\limits_{\rho \to \infty}\frac{\log_2\frac{1}{2}+\log_2\left(\frac{4}{{\rho\bar\kappa}}\right)^{N_r} + \log_2\frac{(2N_r-1)!!}{(2N_r)!!}}{\log_2\rho}.
\end{equation}
Disregarding the terms that are irrelevant to the SNR, we can rewrite (\ref{diver2}) as
\begin{equation}\label{diver3}
\mathcal{D}
=-\lim\limits_{\rho \to \infty}\frac{\log_2\left(\frac{4}{{\rho\bar\kappa}}\right)^{N_r} }{\log_2\rho}.
\end{equation}
Unfolding the numerator term of (\ref{diver3}), we get
\begin{equation}
\mathcal{D}=-\lim\limits_{\rho \to \infty}\frac{2{N_r}-N_r\log_2{\rho}-N_r\log_2{\bar\kappa} }{\log_2\rho}.
\end{equation}
After some arithmetical operations, we obtain the diversity gain as
\begin{equation}\label{diver4}
\mathcal{D}=N_r.
\end{equation}

\subsection{ABEP Expression}
Due to the multiple variables associated with the spatial and symbol domains, it is impossible to obtain the exact ABEP expression directly. Instead, we adopt the the union upper bound of ABEP to evaluate the reliability of proposed scheme as
\begin{small}
\begin{equation}\label{abep}
\begin{aligned}
ABEP \leq &\sum\limits_{n=1}^N\sum\limits_{m=1}^N\sum\limits_{\hat n =1}^N\sum\limits_{\hat m=1}^M\frac{P_b([n,m]\!\to\![\hat{n},\hat m])N([n,m]\!\to\![\hat{n},\hat m])}{MN\log_2MN},
\end{aligned}
\end{equation}
\end{small}%
where $N([n,m]\!\to\![\hat{n},\hat m])$ term denotes the Hamming distance between the transmit signal $h_n$ to $s_m$ and the detected signal $ h_{ \hat n}$ to $s_{ \hat m}$ in the spatial domain and symbol domain, respectively.

\section{Improved TRIS-SM Scheme}
In this section, we enhance the performance of the proposed TRIS-SM system through optimization (\ref{abep}). In essence, the reliability of the constrained proposed scheme depends mainly on the Euclidean distance between neighboring constellation points. As such, optimizing ABEP can be transformed into optimizing the minimum Euclidean distance. Also, this can be corroborated by the CPEP expression in the previous section. Hence, the minimum Euclidean distance becomes a key metric to assess the effectiveness of the proposed system.

Based on (\ref{cpep1}), the square of the minimum receive distance of the proposed scheme can be described as
\begin{equation}
d_{min}^2=\underset{\substack{\substack{\hat m,m \in\left\{1, \cdots, M\right\} \\ \hat n, n \in\left\{1, \cdots, N\right\} \\ \hat m \neq  m \& \hat n \neq n}}}{ \min }\left\|\mathbf{h}_{n}s_m-\mathbf{h}_{\hat n}s_{\hat m}\right\|^2,
\end{equation}
where the $d_{min}$ is a parameter that affects the performance of TRIS-SM scheme, and the system performance can be improved by maximizing the minimum non-zero distance in the constellation.
\subsection{Improved TRIS-SM Scheme }
As stated above, the received minimum distance of the TRIS-SM system can be reformulated as
\begin{equation}\label{adapV01}
\begin{array}{r}
d_{min}
=\underset{\substack{\hat m,m \in\left\{1, \cdots, M\right\} \\ \hat n, n \in\left\{1, \cdots, N\right\} \\ \hat m \neq  m \& \hat n \neq n}}{ \min }\left\|\left[\begin{array}{c}
h_{1, n}s_m  \\
h_{2, n}s_m \\
\vdots \\
h_{N_r, n}s_m
\end{array}\right]-\left[\begin{array}{c}
h_{1, \hat n} s_{\hat m} \\
h_{2, \hat n} s_{\hat m}\\
\vdots \\
h_{N_r, \hat n}s_{\hat m}
\end{array}\right]\right\|_F.
\end{array}
\end{equation}
By the definition of the $F$-norm, we can equivalently transform (\ref{adapV01}) to
\begin{equation}\label{adap01}
\begin{array}{r}
d_{min}
\!=\!\underset{\substack{\substack{\hat m,m \in\left\{1, \cdots, M\right\} \\ \hat n, n \in\left\{1, \cdots, N\right\} \\ \hat m \neq  m \& \hat n \neq n}}}{ \min }\sqrt{
|s_m|^2 \tau_1\!+\!|s_{\hat m}|^2 \tau_2\!-\!2\Re\{s_ms_{\hat m}^*\tau_3\}
},
\end{array}
\end{equation}
where $\tau_1$, $\tau_2$, and $\tau_3$ can be respectively given by
\begin{subequations}
\begin{align}
\tau_1 &= | h_{1, n}|^2+|h_{2, n}|^2+\cdots+|h_{N_r, n}|^2,\\
\tau_2 &= | h_{1, \hat n}|^2+|h_{2, \hat n}|^2+\cdots+|h_{N_r, \hat n}|^2,\\
\tau_3 &=  h_{1, n}h_{1, \hat n}^*+h_{2, n}h_{2, \hat n}^*+\cdots+h_{N_r, n}h_{N_r, \hat n}^*.
\end{align}
\end{subequations}
For the given data rate, we optimize $d_{min}^2$ to improve the reliability of the proposed scheme as
\begin{equation}\label{adap1}
\begin{aligned}
{ \max }  \ \ &d_{min}^2\\
\text { s.t. } &
\frac{1}{N}\sum_{n=1}^{N}\xi_{n}=\bar \xi,
\end{aligned}
\end{equation}
where $\xi_{n}$ denotes the data rate of $n$-th column elements of TRIS, and $\bar \xi$ stands for the average data rate.
\subsection{Simplified Improved (SI) TRIS-SM Scheme }
Due to the very high complexity of (\ref{adap1}), we can simplify the optimization problem by numerical search, given the way the information is transmitted in the TRIS-SM.
To reduce the complexity of (\ref{adap1}), we can make a simplification of it.
Assume that the power of the transmitted signal is normalized, i.e., $|s_m|^2=1$, $\forall m$ and $|s_{\hat m}|^2=1$, $\forall\hat m$.
Considering this, (\ref{adap1}) can be transformed into
\begin{equation}\label{adap12}
\begin{array}{r}
d_{min}^{n,\hat n}
\!=\!\underset{\substack{\hat m,m \in\left\{1, \cdots, M\right\} \\ \hat n, n \in\left\{1, \cdots, N\right\} \\ \hat m \neq  m \& \hat n \neq n}}{ \max }
2\Re\{s_ms_{\hat m}^*\tau_3\}.
\end{array}
\end{equation}
Considering (\ref{adap01}) and (\ref{adap12}), we can simplify (\ref{adap01}) as
\begin{equation}\label{adap2}
\begin{aligned}
{ \max }  \ \ &
\tau_1\!+\tau_2\!-d_{min}^{n,\hat n}
\\
\text { s.t. } &
\sum_{n=1}^{N}\xi_{n}={N}\bar \xi,
\end{aligned}
\end{equation}
where we can calculate the value $d_{min}^{n,\hat n}$ for each candidate.

\begin{remark}
The amplitude of the signal remains constant and fixed as 1 as the symbol domain adopts $M$-ary PSK.
In this case, the improved TRIS-SM and SI-TRIS-SM are the same for both schemes.
\end{remark}

\begin{remark}
Since the simplified and improved TRIS-SM scheme is a process of amplifying the objective function, and because the Q-function is a decreasing function, it can be seen from (\ref{cpep1}) that SI-TRIS-SM scheme is the lower bound of improved TRIS-SM scheme in terms of CPEP.
\end{remark}

\section{Simulation and Analytical Results}
In this section, we validate the ABEP analytical and asymptotic expressions derived in the previous section.
Then, we analyze the ABEP performance of the proposed scheme over the different parameters.
Furthermore, we study the ABEP performance of the improved TRIS-SM scheme.

\subsection{Validation of Theoretical Derivation Results}
\begin{figure}[t]
  \centering
  \includegraphics[width=8cm]{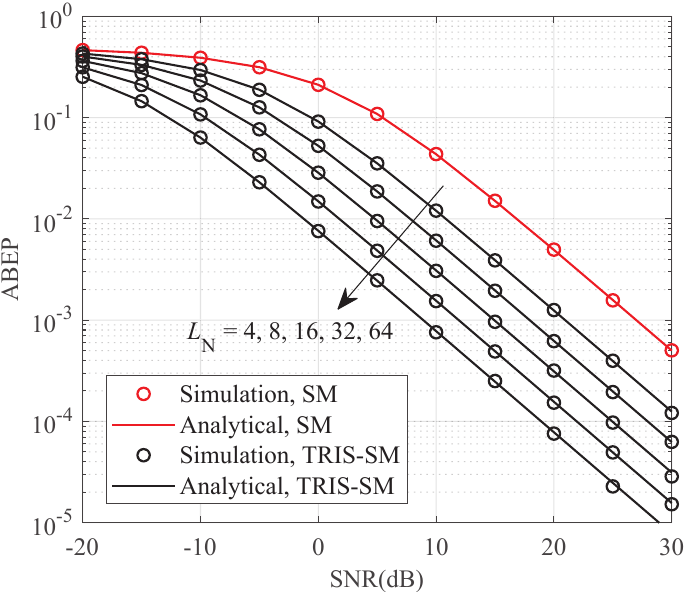}\\
  \caption{\small{ABEP performance comparsion of the TRIS-SM with conventional SM scheme.}}\label{verf1}
\end{figure}

Fig. \ref{verf1} shows the relationship between ABEP and SNR for different numbers of column elements in TRIS. Simultaneously, a comparative analysis of the performance relationship between the proposed TRIS-SM scheme and the conventional SM scheme is presented with the corresponding parameter settings of $M = 1$, $N=2$, and $N_r=1$.
As expected, the analytically derived ABEP agrees well with the Monte Carlo simulation results, where the theoretical curves in this figure are plotted based on (\ref{cpepchar7}).
This phenomenon illustrates the correctness of the theoretical derivation.
It should be noted that (\ref{abep}) represents the union upper bound of the proposed TRIS-SM scheme, where the parameter configuration of this figure is the condition for the equality sign to hold.
According to Fig. \ref{verf1}, it can be seen that when the value of $L_N$ in TRIS is larger, better ABEP performance can be obtained, which is caused by the fact that in the same time slot, when the number of column elements is larger, the energy of the signals that are pooled and transmitted to the receiver will be stronger.
From Fig. \ref{verf1}, it is found that the ABEP value of the proposed TRIS-SM scheme is significantly smaller than the conventional SM scheme. This is because the low-cost nature of TRIS can act as a spatial diversity transmit enhancement for the transmissive signal by increasing the deployment of $L_N$.
\begin{figure}[t]
  \centering
  \includegraphics[width=8cm]{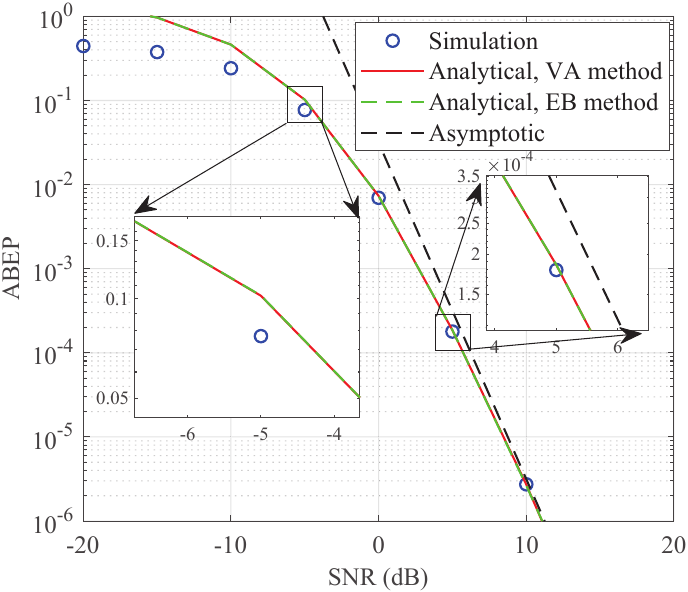}\\
  \caption{\small{Comparative analysis based on  VA and EB methods.}}\label{verf2}
\end{figure}

To evaluate the ABEP theoretical results, we compare and analyse the two methods based on VA and EB in Fig. \ref{verf2}, where the parameters are set to $M=2$, $N = 4$, $N_r = 4$, and $L_N=4$.
It can be observed that the relationship between the simulation results and the analytical curves can be broadly divided into two parts, i.e., the low SNR region and the high SNR region.
It is found that the two match well with each other for SNR greater than -5 dB.
Besides, there is a large gap between the analytical ABEP and the simulation ABEP when SNR is less than -5 dB. The reason for this phenomenon is (\ref{abep}) stands for the union upper bound expression of ABEP for the proposed scheme rather than its exact expression.
Moreover, Fig. \ref{verf2} displays the analytical curves of the VA-based approach and the EB-based two approaches of derived ABEP as the SNR increases. It is clear that the results of the ABEP analytical curves plotted with the two approaches separately are identical over the entire range of SNR. This phenomenon mutually confirms the correctness of the two derived ideas.
Additionally, we also plot the asymptotic results of ABEP in Fig. \ref{verf2}, whereby we find that the asymptotic results match almost perfectly with the simulation and analytical results in the high SNR region. This confirms the accuracy of the asymptotic results and also shows that in the high SNR region we can replace the analytical results with asymptotic in the parameter configuration of this figure.
\begin{figure}[t]
  \centering
  \includegraphics[width=8cm]{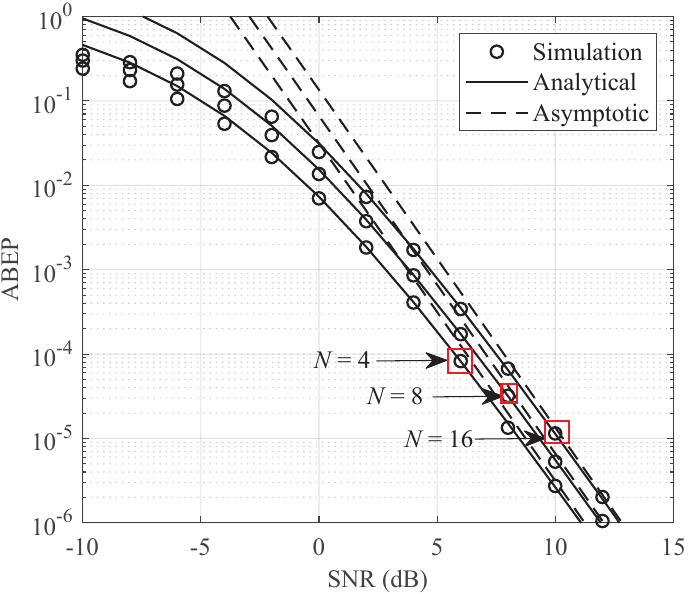}\\
  \caption{\small{ABEP performance  under the different $N$ values.}}\label{N}
\end{figure}
\subsection{Impact of Different Parameters on System Performance}
In Fig. \ref{N}, we depict the relationship of ABEP with SNR for different number of column elements $N$ of TRIS, where the parameters setting in this figure are $M=2$, $N_r = 4$, and $L_N=4$.
The solid and dashed lines represent the theoretical and asymptotic ABEP derived in the previous section, respectively, while the circles with circles represent the ABEP results obtained with Monte Carlo simulations.
As expected, the good agreement between the analytical and simulation results in the middle and high SNR regions validates our derivations. Also,  the asymptotic ABEP gives an upper bound limit for analyzing the ABEP results.
In Fig. \ref{N}, we find that the performance of ABEP decreases significantly as the number of TRIS columns increases, indicating that increasing the number of TRIS in the proposed scheme does not necessarily enhance the performance of ABEP. This is because when the number of TRIS columns increases, it means that the candidate set of the TRIS-SM spatial domain increases. At this time, when the transmit signal power is constant and the probability of selecting elements in each column of TRIS is the same, the possibility of the receiver making an error when making a decoding decision will increase accordingly.

\begin{figure}[t]
  \centering
  \includegraphics[width=8cm]{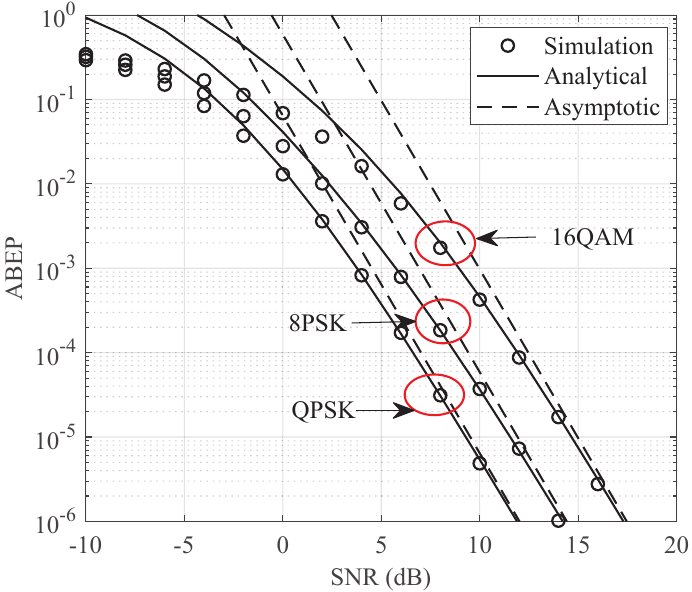}\\
  \caption{\small{ABEP performance under the different $M$ values.}}\label{M}
\end{figure}

Fig. \ref{M} depicts the Monte Carlo simulation, analytical results and asymptotic results of the ABEP versus SNR in dB for different modulation orders. The considered TRIS-SM system employs 4 rows and 4 columns of TRIS and 4 receive antennas.
In this scenario, the simulation results, analysis results, and asymptotic results are represented by circles, curves, and dashed lines in Fig. \ref{M} respectively. At the same time, they are generated respectively corresponding to the equations (\ref{ml}), (\ref{nrpb8}), and (\ref{asyeq4}).
It can be seen that the analytical curve matches well with the Monte Carlo simulation results under high SNR conditions. In addition, as the modulation order is higher, the ABEP performance deteriorates at the same SNR. Specifically, when ABEP = $10^{-6}$, the proposed TRIS-SM system with modulation order QPSK saves 3 dB and 7 dB SNR than the 8PSK and 16QAM-based systems, respectively.
Furthermore, asymptotic results give a quantitative analysis of the trend of the proposed scheme. It is worth mentioning that the analytical curves match the asymptotic results in the high SNR region, which verifies the accuracy of the theoretical derivation, and also indicates that the variation of ABEP tends to be stable with the increase of SNR.

\begin{figure}[t]
  \centering
  \includegraphics[width=8cm]{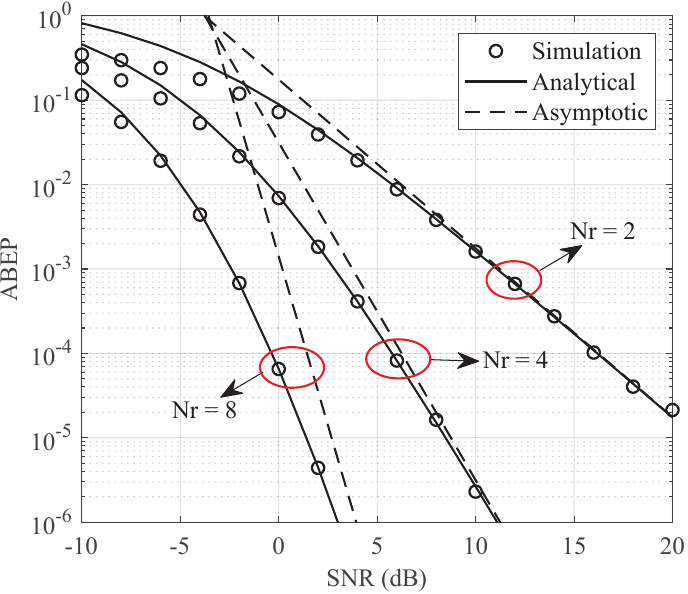}\\
  \caption{\small{ABEP performance under the different $N_r$ values.}}\label{Nr}
\end{figure}

\begin{figure}[t]
  \centering
  \includegraphics[width=8cm]{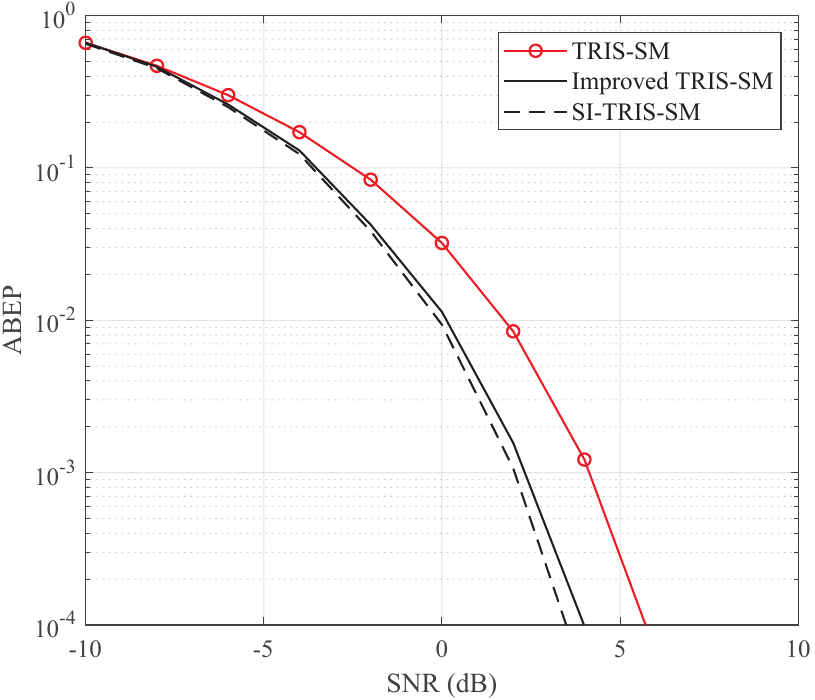}\\
  \caption{\small{{ABEP performance for the improved TRIS-SM scheme.}}}\label{improved}
\end{figure}

In Fig. \ref{Nr}, we investigate the relationship between ABEP and SNR of the proposed TRIS-SM scheme from the perspective of the different numbers of receive antennas, where BPSK modulation is adopted in the symbol domain and $4\times 4$ arrays are adopted in the spatial domain.
From this figure, we observe that the analytical curves based on (\ref{nrpb8}) and the theoretical derivation represented by the curves with Monte Carlo simulation results in circles match perfectly in the medium and high SNR ranges.
Also, we find that the derived asymptotic ABEP results based on (\ref{asyeq4}) and the analytical curves correspond to each other in the high SNR region. This phenomenon not only confirms the correctness of our theoretical derivation but also shows that we can replace the analytical curves with the asymptotic results in the high SNR region to evaluate the performance of the proposed scheme within the error tolerance.
Meanwhile, we also find that more performance gain can be attained by increasing $N_r$ compared to Figs. \ref{N} and \ref{M}.
This is because the transmitter has only one RF link and increasing the number of TRIS rows can only increase the signal transmission area to enhance the reliability of the signal transmission. On the contrary, since each receive antenna is equipped with the corresponding RF link, increasing the number of receive antennas can effectively improve the diversity gain of the system, which is theoretically verified by the (\ref{diver4}).

\subsection{Improved TRIS-SM Scheme Performance}
In Fig. \ref{improved}, we plot the ABEP versus SNR for the improved TRIS-SM scheme, where the parameters $N$, $L_N$, and $N_r$ are set to 2, 4, and 2, respectively.
Since the improved scheme needs to decide on the modulation order based on (\ref{adap1}) the optimal spatial domain and symbol domain modulation orders based on the real-time information of the channel. Then the corresponding modulation is performed, which is an extremely complex process. To simplify this process, we choose the signal transmission rate to be 3 bits per channel use (bpcu) and channel $\mathbf{H}$ as
\begin{equation}
\mathbf{H}=
\begin{bmatrix}
-2.1550 - 1.8483i& -0.2703 + 2.5219i\\
-0.1560 + 2.2516i&  -0.4722 - 1.4695i
\end{bmatrix}.
\end{equation}
In Fig. \ref{improved}, the TRIS-SM scheme uses QPSK, while the improved TRIS-SM scheme uses BPSK and 8QAM under the average 3 bpcu.
As can be seen from Fig. \ref{improved}, the improved TRIS-SM scheme is better than the TRIS-SM scheme in terms of ABEP in high SNR areas.
In addition, we observe that the simplified and improved TRIS-SM achieves the best performance. The reason for this phenomenon is explained in Remark 3.

\section{Conclusions}
In this paper, we proposed a TRIS transmitter-enabled SM-MIMO system scheme that utilizes TRIS column indices to convey data.
At the receiver, the ML detector is used to decode the received signal.
Following the ML detector, we evaluate the ABEP performance of the proposed TRIS-SM scheme. Concretely, we derived the UPEP closed-form expression of the proposed scheme by employing both VB and EB approaches successively. Further, the asymptotic expression of UPEP and the diversity gain of this scheme are provided.  
Furthermore, the joint upper bound expression for ABEP is given by combining the Hamming distance and UPEP results.
In addition, we proposed the improved TRIS-SM scheme to enhance the ABEP performance. To reduce the complexity, the lower bound of the improved scheme is also provided.
Finally, we validated the theoretical analysis results using Monte Carlo simulation. Results reveal that the proposed TRIS-SM scheme is capable of offering better ABEP gain compared to the conventional SM scheme, and the improved TRIS-SM scheme can achieve better ABEP performance.

\begin{appendices}

\end{appendices}

\end{document}